\documentstyle[multicol,prl,aps,epsfig]{revtex}
\begin{document}
\draft
\title{Charge Distribution in a Kondo Correlated Quantum Dot}
\author{D.~Sprinzak, Yang Ji, M.~Heiblum, D.~Mahalu and Hadas~Shtrikman}
\address{Braun Center for Submicron Research, Department of Condensed
 Matter Physics, Weizmann Institute of Science, Rehovot 76100, Israel}
\date{\today}
\maketitle
\begin{abstract}
We report here on a direct and non-invasive measurement of the charge and its distribution in a Kondo correlated quantum dot (QD).  A non-invasive potential-sensitive detector in proximity with the QD reveals that even though the conductance of the QD is significantly enhanced as it enters the Kondo regime the net charge in the QD remains unaffected.  This demonstrates the separation between spin and charge degrees of freedom in the Kondo effect.  We find however, under certain experimental conditions, that an abrupt redistribution of the charge in the QD is taking place simultaneously with the onset of Kondo correlation.  This suggests that the spin-charge separation in the Kondo effect does not always hold.
\end{abstract}


\begin{multicols}{2}
\noindent

\textbf{The Kondo effect, discovered some 70 years ago via an
anomalous behavior of the conductance in metals doped lightly
with magnetic impurities, is one of the most celebrated many-body
phenomena in solid state physics.  The effect results from
many-body interaction between a singly occupied spin-degenerate
level in the magnetic impurity and the electrons in the
reservoirs~\cite{Kondo}.  The many body interaction is believed to lead to a
\textit{spin reconfiguration} in the reservoirs (which screens the
localized spin) without affecting the \textit{charge distribution}
anywhere.  The recent proposals~\cite{Glazman} and subsequent observation of
the effect~\cite{DGG} in a mesoscopic \textit{quantum dot} (QD)~\cite{Meirav} allowed a
systematic study of the effect due to the possibility to vary
many of the system's parameters at will.  Here we report on a
direct, and non-invasive, measurement of the charge distribution
in a QD as Kondo correlation sets in.  A short segment of a
quantum wire (called a \textit{quantum point contact}), located in close
proximity to the QD, served as a potential-sensitive, and hence
charge-sensitive, detector.  The detector revealed that as Kondo
correlation sets in, which is being accompanied by an enhancement
of the QD conductance, the average charge in the dot remained
indeed unaffected.  However, in a certain parameter range, and in
the presence of moderate magnetic field, the onset of the Kondo
effect occurs simultaneously with an abrupt \textit{reconfiguration} of
the charge in the dot.  Moreover, when the correlation diminished
and the QD operated in the ubiquitous Coulomb blockade regime,
charge reconfiguration ceased to take place.  The origin of such
an experimental correlation between charge \textit{reconfiguration} and
spin screening is still unclear, but it certainly deviates from
the orthodox model of the Kondo effect.}

The QD is a small, confined, puddle of electrons weakly connected
to two metallic-like leads (named, \textit{Source} and
\textit{Drain}) with high density of free
electrons~\cite{Meirav}. The confinement leads to an energy
quantization of the electrons (with average level spacing $\Delta$)
and to an additional \textit{charging energy} $U=e^2/C_QD$, which is
required to add an extra electron to the QD ($e$ is the electron
charge and $C_QD$ is the self-capacitance of the QD). Hence, for a
fixed number of electrons in the QD the current at low
temperatures is generally blocked (\textit{Coulomb blockade} (CB) regime),
unless the charging energy is compensated for or the electrons
have a high enough energy.  In the simplest case a QD with an odd
number of electrons behaves as a localized magnetic moment of
spin $1/2$ (due to an unpaired electron), enabling the Kondo
correlation to be established with the free electrons in the
source and drain reservoirs.  Such spin correlation is manifested
via a considerable increase in the conductance of the QD in
regions where otherwise Coulomb blockade would dominate. We
present here a first study of the net charge and its spatial
distribution in a QD tuned to the Kondo correlated regime.  We
probe whether indeed \textit{spin-charge separation} always takes place
via employing a nearby, non invasive, potential detector.  We
show that the separation between spin and charge in a Kondo
correlated system indeed holds under most conditions.  We find,
however, abrupt charge redistribution at the onset of Kondo
correlation under a certain range of experimental conditions.

The experimental configuration, shown in Fig. 1(a), consists of a
small QD, with lithographic diameter of $170 nm$, and a short
segment of a quantum wire (called \textit{quantum point contact},
QPC), electrically separated from it (positioned some $200 nm$
away).  The QD serves as a localized spin coupled to electron
reservoirs and the QPC serves as a potential
detector~\cite{detector}.  Confinement is provided by negatively
biased metallic gates deposited on the surface of a GaAs-AlGaAs
heterojunction embedding a high mobility two dimensional electron
gas (2DEG) some $60 nm$ below the surface.  Gate voltages
$V_{g1}$ and $V_{g2}$ control the coupling of electrons in the QD
to the source ($S_{QD}$) and drain ($D_{QD}$) reservoirs.  As the
coupling becomes stronger (by making $V_{g1}$ and $V_{g2}$ more
positive), spin correlation between the QD and the leads forms,
provided the temperature $T$ is low enough (less than the so
called \textit{Kondo temperature} $T_K$, which represents the
binding energy of the many body spin singlet).  Consequently, the
conductance of the valley of a QD with a net spin increases -
ideally reaching $2e^2/h$ at $T=0$ (so called \textit{unitary
limit}).  The potential within the QPC detector,$\phi_{QPC}$,
depends linearly on the potential within the QD, $\phi_{QD}$,
which in turn depends on the net charge and its distribution in
the QD.  Hence, $\phi_{QPC}=\phi_{QD}\frac{C_{QPC-QD}}{C_{QPC}}$,
with $C_{QPC}$ the self-capacitance of the QPC and $C_{QPC-QD}$
the mutual capacitance between QD and QPC.  The conductance of
the QPC, being initially tuned by $V_g(QPC)$ to
$G_{QPC}=0.5(2e^2/h)$, is highly sensitive to the induced
potential $\phi_{QPC}$.  Running a relatively small current in
the QPC (some $40 nA$), we find that its back-action on the QD is
insignificant.

Let's first understand the expected dependence of $\phi_{QD}$ on
$V_P$ in the CB regime.  As the plunger gate voltage increases
the potential within the QD increases too, reaching eventually
$(e/C_{QD}+\Delta/e$, with $e/C_{QD}$ the potential needed of
charge the QD with one additional electron and $\Delta$ the
quantization energy difference between the single particle
states.  Suppressing fully the charging energy allows an
additional electron to enter the QD and screen the positive built
in potential induced by the plunger gate.  Hence, the potential
within the QD, and consequently the conductance of the QPC
detector, follow a saw-tooth like behavior.  The ideal saw-tooth
shape is usually smeared either by the finite temperature of the
electrons in the leads or by the finite energy width of the
quantized level in the QD,$\Gamma$ (middle part in Fig. 1(b)).
Experimentally, weak features in $\phi_{QPC}$ are more easily
resolved via measuring directly (using lock-in technique) the
differential quantity $dI_{QPC}/dV_P$, which shows a dip instead
of the saw tooth observed in the conductance measurement (bottom
part of Fig. 1(b)).  One of the most useful properties of the
potential detector is demonstrated in Fig. 1(c).  Because of
electrostatic coupling between the plunger gate and the two QD
openings to the leads, highly negative $V_P$ pinches off the QD
inadvertently. Hence, the conductance peaks get weaker as $V_P$
gets smaller and eventually cannot be resolved.  Nevertheless,
the dips in the detector's signal persist further until $V_P=-870
mV$ and cease beyond that voltage, suggesting that the QD is
depleted of electrons.  This allows an exact counting of the
electrons in the QD, as shown in the figure.

We turn now to the Kondo correlated regime, with 9 electrons in
the QD, and focus on the characteristic \linebreak
\begin{figure}
\begin{center}
\leavevmode \epsfxsize=7.5 cm  \epsfbox{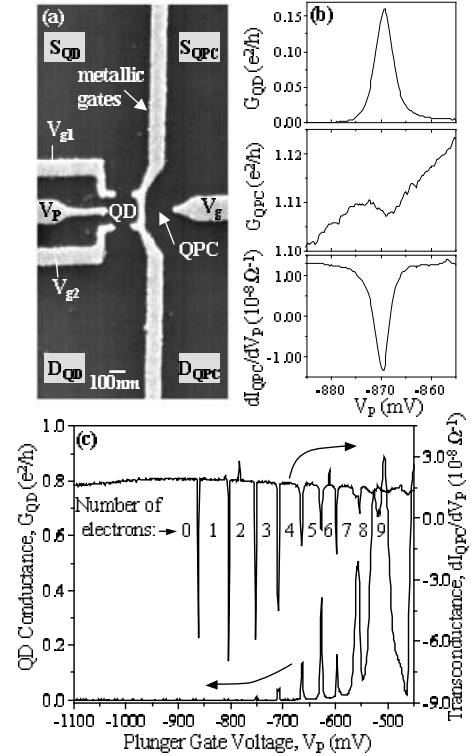}
\end{center}
\vspace{0 cm} \caption{(a) A scanning electron micrograph of the
sample. Negatively biased metallic gates deposited on top of a
GaAs heterostructure containing a 2DEG below the surface (with
electron density, $n=3.1\times10^{11} cm^{-2}$ and mobility
$\mu=5\times10^5 cm^2V^{-1}s^{-1}$), forming a small QD (on the
left) and a QPC detector (on the right).  The QD configuration is
tuned via the Plunger Gate Voltage, $V_P$ and by $V_{g1}$ and
$V_{g2}$ that control the coupling to the leads.  Both the
conductance through the QD and the detector signal are measured
simultaneously.  (b) Top--- a typical CB peak in the QD
Conductance measurement. Middle --- A typical conductance
measurement through the QPC detector showing the saw-tooth like
shape due to addition of one electron. The QPC is tuned to a
transmission of about 0.5 of the first conducting channel
($G_{QPC}\sim e^2/h$). This conductance is proportional to the
average electrostatic potential in the QD,$\phi_{QD}$ . Bottom
--- a differential measurement of the detector signal.  The
measurement is performed by adding a small AC modulation to $V_P$
(1 mV rms in this case) and measuring the current modulation
through the QPC, $I_{QPC}$ (1 mV DC bias is applied between the
source and the drain of the QPC).  This technique provides a much
cleaner detector. (c)  Typical measurement of the QD conductance
(bottom curve) and the detector signal (top curve) as a function
$V_P$. Each peak in the conductance has a corresponding dip in
the detector.  The regime of $V_P < -870 mV $ corresponds to zero
electrons in the QD (see text). The number of electrons in each
CB regime is shown on the figure.  The double peak corresponding
to 9 electron is in the Kondo regime. }
\end{figure}

\begin{figure}
\begin{center}
\leavevmode \epsfxsize=8.5 cm  \epsfbox{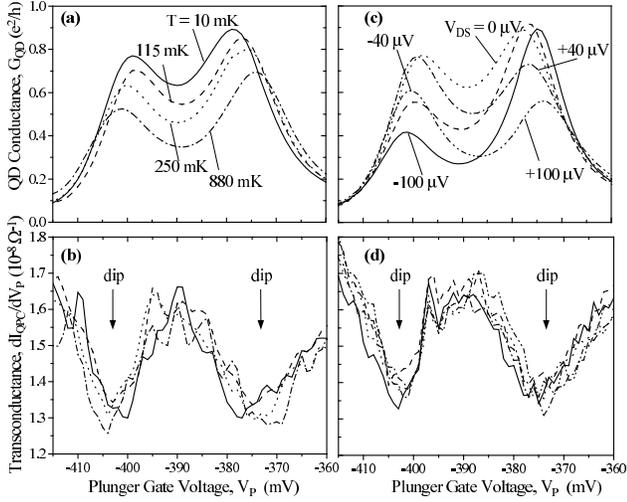}
\end{center}
\vspace{0 cm} \caption{(a) Conductance through the QD as a
function of $V_P$, for different temperatures (the scan is over
the region of VP where $9 +- 1$ electrons are in the QD but with
a different QD configuration than in Fig. 1(c). The observed
increase in conductance as temperature decreases is a
manifestation of the Kondo correlation.  (b)  Detector signal
measured simultaneously with QD conductance in (a). It is clear
that while the conductance is varied considerably with
temperature, due to Kondo correlation, the detector signal and
hence the net charge in the QD are unchanged.  (c) and (d)
Conductance through the QD and detector signal as a function of
$V_P$ for different Drain-Source bias on the QD, $V_DS$.  While
the conductance is strongly quenched by applying either positive
or negative $V_DS$, the detector signal, and hence the net charge
in the QD, are unchanged (the asymmetry in the conductance
between positive and negative bias is a result of an asymmetric
QD configuration). }
\end{figure}
\noindent
double peak in the conductance(see Fig. 1(c)).  A first
question to ask is whether the net charge in the QD is conserved
when Kondo correlation is being established via decreasing the
temperature or lowering the source -- drain bias across the QD
--- both leading to an enhancement in the valley conductance.
The detector signal in Fig. 2, for different temperatures and
different source – drain biases indeed show that in both cases
the potential evolution, hence the average net charge in the QD,
do not change significantly as Kondo correlation is established.
This leads to the conclusion that the enhancement in the valley
conductance (which is due to the Kondo correlation) results from
a larger number of electrons traversing the QD, each dwelling for
a much shorter time.  This short dwell time is justified by the
virtual nature of the co-tunneling process in the Kondo
effect~\cite{Meir1}.  This is a clear and direct proof that spin
correlation is established without affecting the average charge
distribution in the system.~\cite{PC1}

A careful examination of the detector signal in Fig. 1(c) reveals
at least one sharp dip superimposed on the\linebreak
\begin{figure}
\begin{center}
\leavevmode \epsfxsize=8.5 cm  \epsfbox{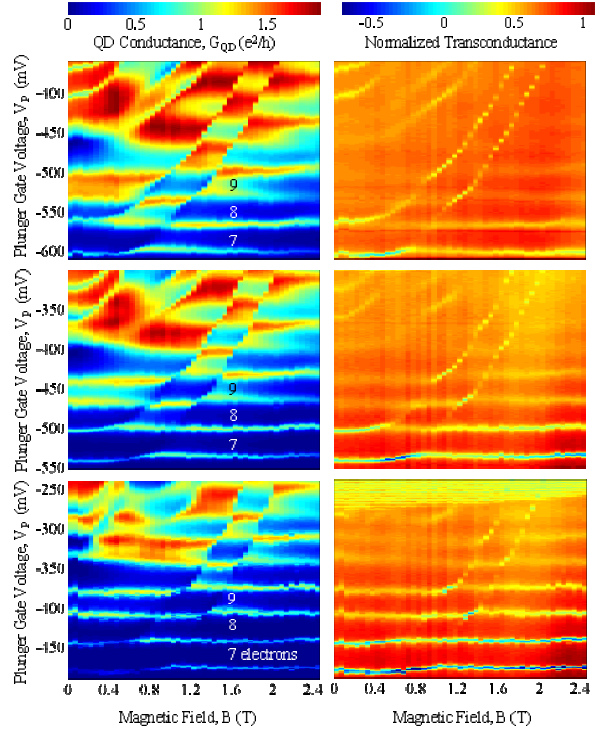}
\end{center}
\vspace{0 cm} \caption{2D color plots of the conductance through
the QD (left plots) and the detector signal (right plots) as a
function of $V_P$ and magnetic field perpendicular to the device,
$B$.  Top --- strongest coupling to the leads, bottom --- weakest
coupling to the leads. The three different coupling strengths are
obtained by changing the two gate voltages – $V_{g1}$ and
$V_{g2}$ (more negative bias for weaker coupling) and
compensating by the bias on VP so that the QD is in the same
electronic configuration.  Number of electrons is marked for some
CB valleys on the conductance plots.  The color scale
representation is given at the top of the figure.  The detector
signal in each plot is normalized by the maximum value of each
plot since the detector signal sensitivity depends on the QD
configuration.  The horizontal lines in the figures correspond to
CB peaks separating regions with the same number of electrons.
Another clear set of lines curving upwards are boundaries that
delineate a chessboard like pattern of high (red/yellow) and low
(blue) conductance in the conductance plots.  In the detector
signal these lines are essentially sharp dips corresponding to
charge reconfiguration occurring in the QD (see text). }
\end{figure}
\noindent
broader, main dip, in the transition region between 7
and 8 electrons in the QD.  The emergence of such dips is made
clearer in a two dimensional (2D) color plots of the detector
signal and conductance as a function of plunger gate voltage,
$V_P$, and magnetic field, $B$, applied in a direction normal to
the plane. Both, 2D plots of QD conductance and detector signal
reveal rich, \textit{chessboard-like} spectra (see Fig. 3(a)).
The main features in both plots are the horizontal yellow/blue
lines, which represent the familiar CB peaks and the corresponding
detector dips (the derivative of the saw tooth like potential due
to a single electron entering the QD).  Being independent of
magnetic field, these horizontal lines are indicative of the
relative insensitivity of electron filling to magnetic field in
the scanned range.  Other prominent features in the 2D plot of
the detector signal are the sharp dips, being initially
superimposed on the broader dips and departing from them at
higher magnetic fields to curve upwards and cross the broader
horizontal lines.  In exact coincidence with the detector dips,
the 2D plot of the conductance exhibits distinct boundaries
separating regions of high (red/yellow) and low (blue)
conductance.  The regions of high valley conductance are
identified as Kondo enhanced valleys while those of low
conductance as CB valleys (such behavior was also recently
observed by J. Schmid \textit{et al.}~\cite{Schmid}).  The sharp
dips in the detector signal result from abrupt changes in the
potential of the QD, which in turn must result from abrupt
rearrangements in the charge of the QD (while the total net
charge in the QD remains constant)~\cite{PC2}.  This coincidence
between the detector dips and the onset of Kondo correlation
surprisingly connects charge reconfiguration in the QD and the
spin correlation of the Kondo effect.

Pinching off the QD, by applying negative bias voltage to gates
$g_1$ and $g_2$, or indirectly via negatively biasing the plunger
gate, confines the electrons to the inner part of the QD and
decouples the dot from the reservoirs.  This leads to quenching
of the Kondo correlation at finite temperatures.  As seen in
Figs. 3(b) and 3(c), for weaker coupling to the leads the valley
conductance quenches while, simultaneously, the detector dips
become smaller.  For strongly pinched off QD the detector dips
disappear altogether.  A similar correspondence can be seen in
Fig. 4 as the temperature increases. Both QD conductance and
detector's sharp dips quench approximately with the scale of the
so-called Kondo temperature, $T_K$ (the approximate binding
energy of the many body spin singlet).  Note that even though the
application of VDS quenches the valley conductance, Kondo
correlation with electrons in each lead separately persist.  Such
an experiment indeed result (results not shown) with the
splitting of the dips to doublets.

The peculiar coincidence  between the abrupt charge
reconfiguration (the sharp dips in the detector) and the onset of
Kondo correlation (chessboard like pattern in conductance) can be
explained if one takes into account level crossing in the QD
spectrum.  Namely, as B and Vp vary, quantum states can
(energetically) cross each other and charge can be transferred
from one state to another, leading to a spatial redistribution of
the charge accompanied with a spin reconfiguration in the QD.  We
propose here a model~\cite{PC3}, depicted in Fig. 5, based, in
principle, on the energy level spectrum of a two-dimensional
parabolic well with a magnetic field threading it
perpendicularly~\cite{Darwin}.\linebreak
\begin{figure}
\begin{center}
\leavevmode \epsfxsize=5.5 cm  \epsfbox{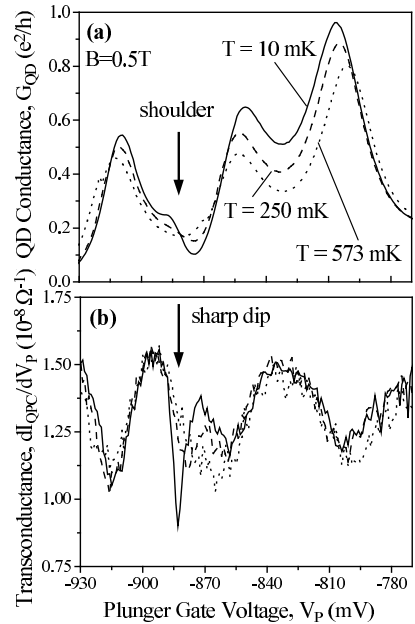}
\end{center}
\vspace{0 cm} \caption{Temperature dependence of the QD
conductance and the 'sharp dip' in the detector signal. The sharp
dip in the detector signal (b), corresponding to charge
reconfiguration in the QD, smears as the temperature increases on
the scale of $T_K$.  The measurement is performed at $B=0.5 T$ so
that the sharp dip appears approximately in the middle of a CB
valley. The visible shoulder in the conductance forms because a
transition from a Kondo regime (left of the shoulder) to a
non-Kondo regime (right of the shoulder) occurs at this point. }
\end{figure}
\noindent
We assume in our model that some states that are
localized to the center of the QD (hence having weak coupling to
the leads) have strong magnetic field
dependence~\cite{Comment1}.  Consequently, by increasing magnetic
field, these states cross more extended states (that are better
coupled to the leads), which have weaker magnetic field
dependence.  Such crossings, as is shown next, can lead to the
basic phenomena observed --- charge reconfiguration and onset of
Kondo correlation.  We neglect Zeeman splitting in our model
since it quenches the Kondo effect only at a higher magnetic
field range (above $2-3T$)~\cite{DGG}.

We now apply these assumptions  to our experimental results to
see how the model works.  We first schematically reproduce the 2D
plot of Fig. 3 in Fig. 5(a), with \textbf{K} for Kondo regimes
and \textbf{NK} for non-Kondo regimes.  We start first with
constant $V_P=V_{P1}$, namely, a spin polarized QD with seven
electrons.  Since at $B=0$ we are at a NK regime we assume that
the seventh electron occupies a localized state (with weak
coupling to the leads).  The single particle spectrum ($E-B$
diagram) that corresponds to this case is shown in Fig. 5(b).  As
the magnetic field increases the energy of the localized state
increases, crossing eventually that of the extended\linebreak
\begin{figure}
\begin{center}
\leavevmode \epsfxsize=8.5 cm  \epsfbox{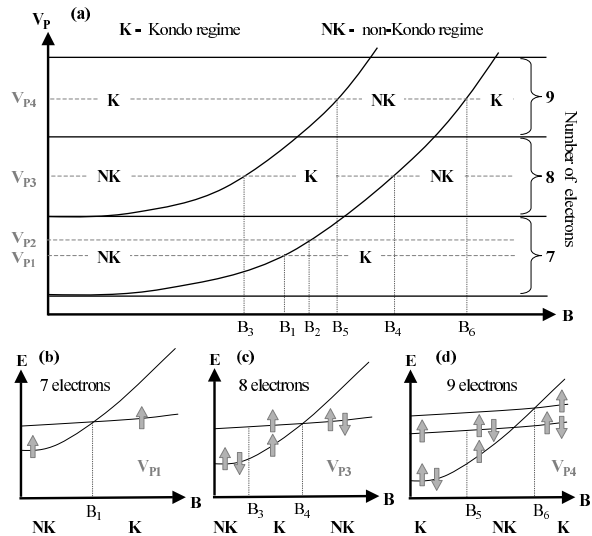}
\end{center}
\vspace{0 cm} \caption{(a)  2D $B-V_P$ diagram reconstructing the
bottom half of the top plots in Fig. 3. Horizontal solid lines
correspond to CB peaks (separating regions with different
electron number).  Curved solid lines correspond to the charge
reconfiguration lines (i. e. sharp dips in the detector).  The
crossing of these sets of lines forms the 'chessboard pattern'.
We mark the regions with Kondo correlation by K and the ones
without Kondo correlation by NK.  Gray dotted lines (horizontal)
marked $V_{P1}$ to $V_{P4}$ are constant $V_P$ lines referred to
in the text. (b), (c) and (d) are the suggested QD spectrum of
levels in our model as a function of magnetic field corresponding
to $V_{P1}$, $V_{P3}$, and $V_{P4}$ ,respectively.  We assume
that the level with the strong magnetic field dependence is more
localized to the center of the QD and hence has a weaker coupling
to the leads.}
\end{figure}
\noindent
state at $B=B_1$.  Beyond the crossing point the seventh
electron, seeking the lowest energy crosses to a more extended
state that is better coupled to the leads.  Following this
crossing the detector produces a sharp dip in the detector (since
the charge of the electron was redistributed~\cite{Comment2}) and
a Kondo cloud is formed (since the electron is better coupled to
the leads). Increasing $V_P$ to $V_{P2}$ (still maintaining seven
electrons in the QD, Fig. 5(a)) lowers the energy of the
localized state faster than that of the extended
state~\cite{Comment3}, thus moving the crossover energy of both
states to a higher field $B=B_2$ (in Fig. 5(a)).  This reproduces
the 2D path of the sharp dip for seven electrons in the QD.

Increasing the plunger gate voltage further to $V_P=V_{P3}$ (in Fig. 5(a)) increases the number of electrons in the QD to eight.  The last two electrons occupy the same localized state and the QD has no net spin.  As the magnetic field increases the localized state approaches the extended state in Fig. 5(c) and at $B=B_3$it becomes energetically favorable for one electron to flip its spin and jump to the extended state.  This way total energy is minimized since electron-electron interaction energy (being large for two electrons in a localized state) is reduced by such a flip.  This process is similar to a singlet-to-triplet transition already observed in QDs~\cite{Kouwenhoven} and to the recent observation of the Kondo effect in the unitary limit, which was attributed to such a transition~\cite{Wiel}.  The transition reconfigures the charge and polarizes the QD, allowing Kondo correlation to take place.  At a higher magnetic field, $B=B_4$, level crossing takes place and it becomes energetically favorable for the last two electrons to occupy the extended state with net spin zero.  Consequently, the charge reorganizes again and Kondo correlation ceases.  Similar arguments apply when $V_P=V_{P4}$ with nine electrons in the QD and the ninth electron is in an extended state at $B=0$ leading to a Kondo regime, as depicted in Figs. 5(a) and 5(d)~\cite{Comment4}.

This model explains the main features of our experiment and the correlation between the onset of Kondo effect and charge reconfiguration.  It is not clear from it why when Kondo correlation is quenched (via higher temperatures or smaller coupling to the leads) the detector dips also disappear.  Since level crossing and charge redistribution is expected also to take place in a QD in the CB regime, our detector suggested that it is extremely small and cannot be measured.  One may attribute this effect to the smaller difference between the capacitance of the localized state and the extended state when the QD is weakly coupled to the leads (namely, in the CB regime).  Alternatively, the strong correlation between the appearance of charge reconfiguration and the onset of the Kondo effect may suggest \textit{spin-charge correlation} – as the electron moves from a localized to an extended states (or back).  Nevertheless, in the absence of magnetic field, when charge reconfiguration is not found, we show that Kondo correlation is established without affecting the average net charge in the QD - demonstrating \textit{spin-charge separation} in the Kondo correlated regime.


\textbf{Acknowledgement :}  We thank Y. Oreg, Y. Meir, A. Silva and S. Levit for many fruitful discussions. The work was partly supported by the MINERVA foundation and the German Israeli Project Cooperation (DIP).
Correspondence should be addressed to M.H. (email: heiblum@wisemail.weizmann.ac.il)

%
%

%
%
\end{multicols}

\begin{references}
\bibitem{Kondo} Kondo,~J., in Solid State Physics, Eds. H.~Ehrenreicht,
F.~Seitz, D.~Turnbull, Academic Press, New York, \textbf{Vol. 23}, 13~(1969);
Anderson,~P.~W.  Localized magnetic states in metals. Phys. Rev. \textbf{124}, 41~(1961).
\bibitem{Glazman} Glazman,~L.~I. and Raikh,~M.~E. Resonant Kondo transparency of a barrier with quasilocal impurity states. JETP Lett. \textbf{47}, 452~(1988);
T.~K.~Ng and P.~A.~Lee,. On-site Coulomb repulsion and resonant tunneling.  Phys. Rev. Lett. \textbf{61}, 1768~(1988).
\bibitem{DGG}Goldhaber-Gordon,~D.,Shtrikman,~ H.,
Mahalu,~D.,Abusch-Magder,~ D.,Meirav,~ U.,Kastner,~ M.~A. Kondo effect in a single electron transistor. Nature \textbf{391}, 156~(1998).
\bibitem{Meirav}Meirav,~U.  and Foxman,~E.~ B. Single-electron phenomena in semiconductors. Semicond. Sci. Technol. \textbf{10}, 255~(1995).
\bibitem{detector} Field,~M. \textit{et al.} Measurements of Coulomb blockade with a noninvasive voltage probe. Phys. Rev. Lett. \textbf{70}, 1311 (1993); Buks,~E., Schuster,~R., Heiblum, ~M., Mahalu,~D. and Umansky,~V. Dephasing in electron interference by a 'which-path' detector. Nature \textbf{391}, 871~(1998); Berman,~D., Zhitenev,~N.~B., Ashoori,~R.~C. and Shayegan,~M. Observation of quantum fluctuations of charge on a quantum dot. Phys. Rev. Lett. \textbf{82}, 161~(1999).
\bibitem{Meir1}  Meir,~Y., Wingreen,~N.~S. and Lee,~P.~A. Low-temperature transport through a quantum dot: The Anderson model out of equilibrium. Phys. Rev. Lett. \textbf{70}, 2601~(1993); Wingreen,~N.~S. and Meir,~Y. Anderson model out of equilibrium: Noncrossing-approximation approach to transport through a quantum dot. Phys. Rev. \textbf{B49}, 11040~(1994)
\bibitem{PC1} The correction to the average charge is expected to be negligibly small (Meir~Y., Private communication).
\bibitem{Schmid}Schmid,~J., Weis,~J., Eberl,~K. and Klitzing,~K.~v. Absence of odd-even parity behavior for Kondo resonances in quantum dots. Phys. Rev. Lett. \textbf{84}, 5824~(2000).
\bibitem{PC2} Since charge is not exactly integer due to charge fluctuation to the leads there could still be some small change in total average charge by such transition (Meir,~Y. and Wingreen,~N., private communication).
\bibitem{PC3} Oreg,~Y. , Meir,~Y. and Wingreen,~N., private communication
\bibitem{Darwin} Darwin,~C.~G. The diamagnetism of the free electron.  Proc. Cambridge Philos. Soc.Math. Phys. Sci.\textbf{27}, 86~(1930); Fock,~V. Bemerkung zur quantelung des harmonischen oszillators im magnetfeld, Z. Phys. \textbf{47}, 446~(1928).
\bibitem{Comment1} One can show in the Darwin-Fock spectrum11 that these states that cross other states by increasing magnetic field (like shown schematically in Fig. 5(b)) are in most cases more localized to the center of the QD than the states being crossed (with a few electrons in the QD).
\bibitem{Comment2} The effect of the detector can be understood in terms of a change in QD capacitance. Assuming this capacitance is different between localized and extended states, the jump in the QD potential is evident from the formula   , where CP-QD is the capacitance between the plunger gate and the QD.
\bibitem{Comment3}  The rate at which a level shifts with VP depends on the QD capacitance - as seen in the formula in reference~13.
\bibitem{Kouwenhoven} Kouwenhoven,~L.~P.\textit{et al.} Excitation spectra of circular, few-electron quantum    dots. Science \textbf{278}, 1788~(1997).
\bibitem{Wiel} van~der~Wiel,~W.~G. \textit{et al.} The Kondo effect in the unitary limit. Science \textbf{289}, 2105~(2000).  Nygard, ~J., Cobden, ~D.~H. and Lindelof,~P.~E., Kondo physics in carbon nanotubes. Nature \textbf{408}, 342~(2000).
\bibitem{Comment4} In this case one has to note that electron-electron interaction is stronger in the localized level (that's why an electron still jumps to the next level) and that one more strongly coupled level has to be involved

\end{references}
\end{document}